# Enabling Big Data Analytics at Manufacturing Fields of Farplas Automotive

Özgün Akın[1], Halil Faruk Deniz[1], Doğukan Nefis[1], Alp Kızıltan[1] and Altan Çakır[2]

[1] Farplas Otomotiv, 41400 Kocaeli, Turkey
[2] Istanbul Technical University, 34467 Istanbul, Turkey
o.akin@farplas.com

**Abstract.** Digitization and data-driven manufacturing process is needed for today's industry. The term Industry 4.0 stands for the today industrial digitization which is defined as a new level of organization and control over the entire value chain of the life cycle of products; it is geared towards increasingly individualized customer's high-quality expectations. However, due to the increase in the number of connected devices and the variety of data, it has become difficult to store and analyze data with conventional systems. The motivation of this paper is to provide an overview of understanding of the big data pipeline, providing a real-time on-premise data acquisition, data compression, data storage and processing with Apache Kafka and Apache Spark implementation on Apache Hadoop cluster, and identifying the challenges and issues occurring with implementation the Farplas manufacturing company, which is one of the biggest Tier 1 automotive suppliers in Turkey, to study the new trends and streams related to topics via Industry 4.0.

**Keywords:** Big Data, Apache Hadoop, Apache Kafka, Apache Spark, Apache Ni-Fi, AI in Manufacturing, Industry 4.0, MQTT, OPC

## 1 Introduction

### 1.1 Big Data in Manufacturing

In today's modern world, companies need to grow in an increasingly competitive and changing environment. In this manner**,** knowledge is a major source of competitive advantage and value creation to the development of dynamic capabilities, and this is reflected in organizational success. The last decade has witnessed massive progress in the enterprises' revenues that employed big data in their businesses. Accordingly, most business activities take place in a big data environment, and in fact, big data are required for many companies to survive [1].

Injection molding is a manufacturing process for producing plastic parts by injecting plastic raw material into a mold. A wide variety of products are produced using injection molding, which varies greatly in size, complexity and applications. According to Chen, Nguyen, and Tai mentioned in [2] that there are two most common problems in injection molding. The first problem is to derive the optimal process parameters for the initial injection molding machine setting, considering several process variables to implement effective manufacturing after product design. To derive optimal process variables, there is a process that causes significant losses to constantly adjust process parameters in trial and error mode. The second problem is the difficulty in changing the value of the initial process parameter during the injection molding process, especially when defects occur during manufacturing. Currently, there is no method for quality enhancement of this dilemma yet, and it is difficult to react to equipment failure and other unexpected production problems. In addition, a fuzzy-based controlled classifier can be used in the automotive to search, store and classify Big data.

Farplas is an automotive parts supplier company founded in 1968. The company has been able to catch up with the technological trends and adopted the vision of offering innovative solutions to the mobility ecosystem. Due to high-quality standards of the company, a big data analytics project has been started recently. In the following sections, the Farplas big data project pipeline is summarized: data ingestion, steps to get data from the production line to a big data lake, data transformation and storage, the method used to transform and store the data to prepare analytics and visualization, data analytics, to extract meaningful things from this data, and data visualization, to visualize these results.

## 1.2 Data Ingestion

The data collection phase is the first step to be achieved towards developing data-based decision mechanisms and artificial intelligence-supported production systems. Collecting data in various sizes and frequencies from the production points is of great importance for more meaningful analysis [3]. Since the data sources in production are not uniform, many difficulties are encountered during the data collection phase.

More sensors are involved, especially in the industrial field, to obtain smart products, manufacturing equipment and production, big data applications have become an essential element of the operation in the period of Industry 4.0 [4]. According to data-based manufacturing, Farplas has deployed a pilot injection cell to test challenges for intelligent production. Thus, the most suitable system solution is studied in the pilot cell and the best solution would easily spread to all injection molding machines of the company. Sensors are placed in various areas of the pilot injection machine at Farplas. It is aimed to transfer the data created in the system to the big data platform, which will be called as Flatform instantly.

The pilot injection cell at Farplas consists of the following seven components for the analysis: a plastic injection molding machine, 6 axes robot, energy analyzer, water collector, precision scale, ambient humidity, and temperature sensor. All these devices use different protocols to communicate with systems, for example, Euromap63, which is a corresponding protocol specific to plastic injection machines is used to get data from plastic injection machines [5]. What's more, the main obstacle of this protocol is to make the data meaningful using complex parameter names, which are for different brands and models of machines. The data received with the Euromap63 sensor is carried to the main PLC board via ethernet. The energy analyzer, water collector and ambient sensor are connected to the main PLC and parameters are defined using machine name

4and model in OPC Software (Kepware). Afterwards, these parameters are published from Kepware over MQTT protocol in Fig. 1.

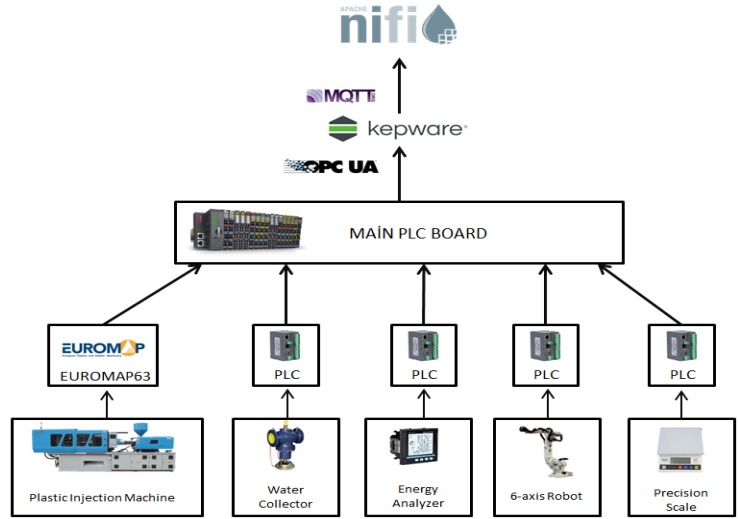

**Fig. 1.** The flow used for data ingestion in Farplas.

Subsequently, the data is transferred to the Flatform according to a certain rule. At this point, Apache Ni-Fi is used in the Flatform to enable data transfer and create data pipelines [6]. Ni-Fi, which consumes data with the Consume-MQTT processor, is preferred to create all data pipeline in the Flatform. The conversion is made in JSON format consumed by Ni-Fi. The transformed data is transferred to Kafka and the Kafka distributes the data to HDFS and Elasticsearch and distributes the data to other platforms.

### 1.3 Data Transformation and Storage

After data collection, the raw data must be transformed in data storage platforms due to the huge volume, redundancy, vulnerability highlights of data structures. In this respect, the application of data transformation starts with data cleaning, data integration and data compression [7].

According to [8], Ni-Fi is utilized for integrating data from different data sources and gives real-time control that performs ETL basic to oversee the movement of data between sources and destinations. Using Ni-Fi, the data mobility, from HDFS and Elasticsearch, is consistently integrated with up-to-date tools. Ni-Fi is data source with supporting disparate and distributed sources of translating varying formats, schemas, protocols and sizes such as machines, geo-location devices, files, log records and among others [9]. Ni-Fi has several processors operated to data extraction, transformation and loading (ETL), making available connectors for the file system in the Hadoop Cluster, Elasticsearch and distributed messaging technologies such as Kafka.

Based on our prerequisites, proposed Apache Kafka, Apache Spark on Apache Hadoop and Elasticsearch combination is the foremost appropriate big data pipeline. The proposed system is implemented to monitor the injection molding process in real-time as shown in Fig. 2. In this manner, data streaming and transmission are essentially required in the proposed system. Following the results of the recent paper [10], it is accepted that Spark and Kafka could be one of the best choices in terms of real-time data streaming and transmitting. All this installation environment is coordinated to work together, like a Hadoop ecosystem together with an *In-Memory-Flow* approach with Apache Spark.

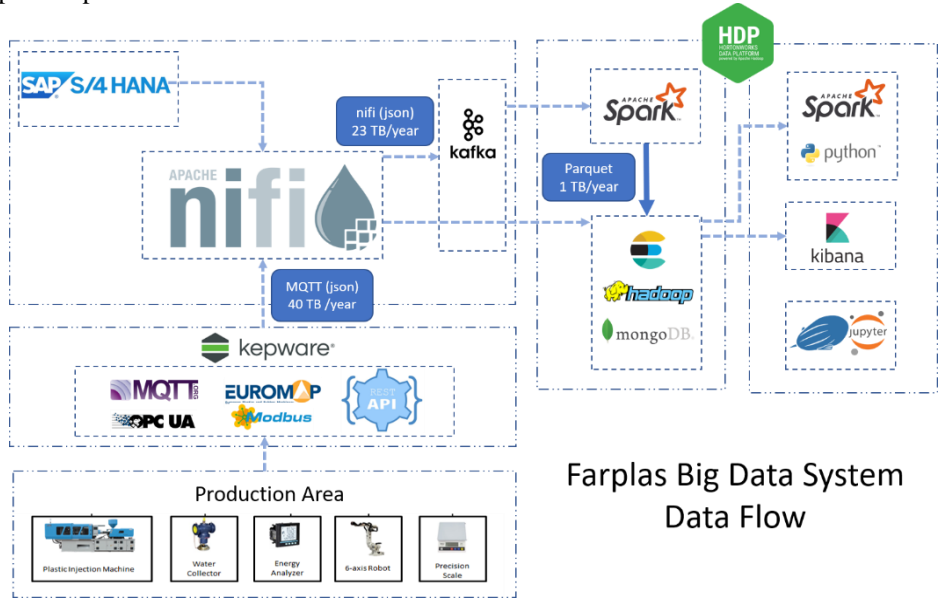

**Fig. 2.** Real-time distributed Big Data architecture and parquet transformation of raw data used in Flatform.

Within the data storage layer, the large volume of collected and processed data from the production line must be securely stored and integrated consistently. Furthermore, mass productional data beneath the on-premise environment have high prerequisites for data storage and preprocessing. Moreover, Hadoop cluster can support the development and accomplish the parallel processing of large-scale data. As one primary way to accomplish on-premise computing, Hadoop can build large-scale cluster framework on common hardware and has advantages of low-cost, extensible, proficient and dependable [11].

According to our tests in Farplas, about 40 TB data will be generated annually. For this reason, data size can be a major concern. Therefore, to reduce the estimated size of the data, Apache Parquet, is utilized. As seen in Fig. 2, Parquet is a machine-readable columnar storage format available in the Spark+Hadoop ecosystem, which is strongly supported by Spark, and it provides around 10-30 times more compressed than JSON format.

### 1.4 Data Analytics

The point to be reached with the analytics layer is creating data-driven manufacturing to make financial and soft gains. All data stored in the data lake and it is reachable from the analytics layer in the Flatform. Analytics layer contains IDE applications, which are Jupiter and Zeppelin to manage Spark and Python codes to build real-time data processing and machine/deep learning models.

The output of the model is visualized in Kibana by sending it to the Elasticsearch database via spark as often as desired. Farplas uses discrete manufacturing method with over 80 injection molding cells and produces approximately 70 million parts per year. Furthermore, some of the produced parts take the final shape after having assembly and painting processes. The surface quality of the parts is very important in terms of individual cost and company prestige. Therefore, the predictive quality model is selected as the first use-case scenario.

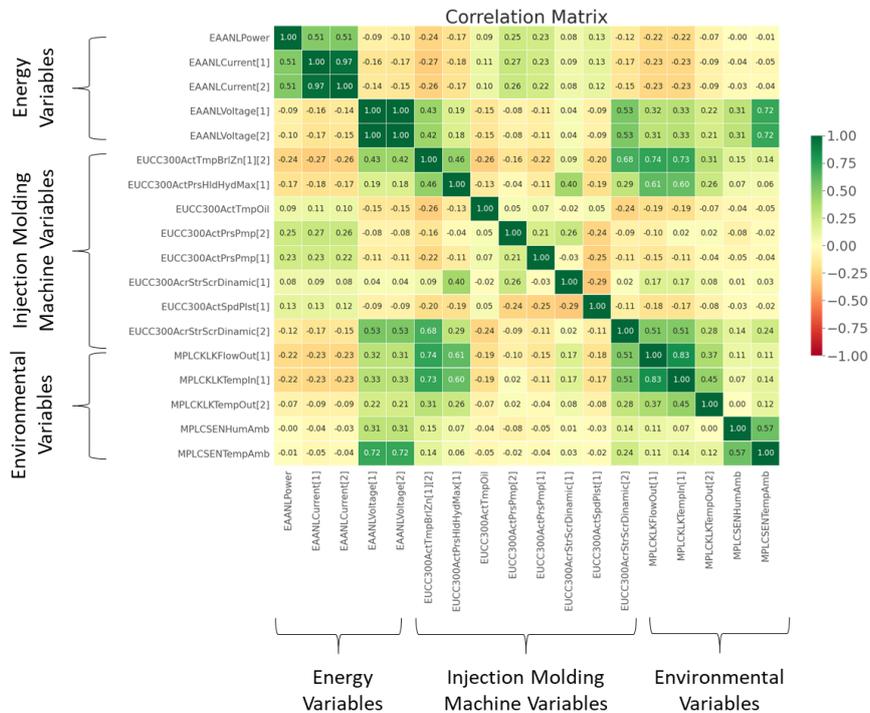

**Fig. 3.** Correlation matrix between machine variables and sensor variables.

The first approach for the predictive quality application is the correlation matrix method to examine production data and relations between parameters in real-time observation, as seen in Fig. 3. Thanks to the correlation matrix relevance between parameters which are produced in the cycle are visualized for each production cycle. The expected situation is that the correlation matrix of each cycle is highly like the others.

Nonetheless, it is seen that some of the cycles have a different correlation matrix when compared to the others. These cycles are easily labelled as a problematic cycle.

The second approach, each cycle is represented as vectors consisting of process data collected for itself. Then, the distances of each cycle vector with the 10 vectors before that cycle are calculated. When the similarities are displayed with colors, loops with abnormalities are detected. Compared to the correlation application performed above, it is not clearly shown which parameter the problem originates from, but since the analyzed data is more, the detection accuracy rate of the application is higher.

The second application which is selected to detect anomalies in the production and predict the quality of the part is the cosine similarity integration with collaborative filtering. In this real-time data analysis approach, each vector represents the process data collected during the cycle time. After that, the distances of each cycle vector with the 10 cycle vectors before that cycle are calculated. When the similarities visualized with colors, cycles that have anomaly are detected. However, this approach has not clearly shown which parameter has the error yet.

In the third application, the correlation matrix application is combined with the cosine similarity application and converted into a cosine matrix. In the cosine matrix, the parameter values produced in each second for each machine cycle are defined as vectors, and the distance of the parameter vectors within that cycle is found by the cosine similarity method and converted into a matrix and colored. It is seen that when the color distributions on the matrices of each cycle are examined, the color distributions of the abnormal cycles are different from the others.

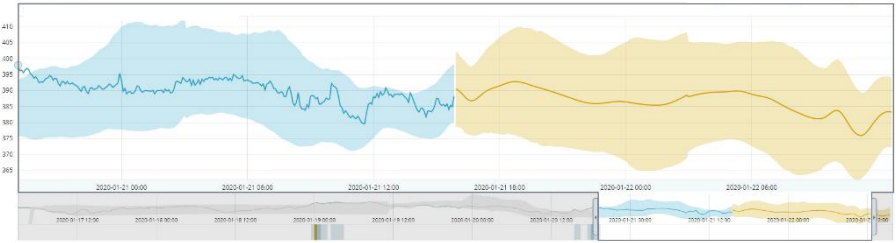

**Fig. 4.** Time series analysis of forecasting energy consumption for the next consecutive cycles.

In addition to the predictive quality application, another application that will enable us to gain data is energy analysis. Based on the energy values consumed by the machines, abnormalities in production can be detected, and optimization efforts can be made to reduce energy consumption, thereby saving money from energy consumption. The first application in Flatform is by collecting the energy data of the machine and analyzing it with the machine learning time series methods. The abnormalities formed by detecting the data outside these confidence intervals (see Fig. 4).

### 1.5 Data Visualization

In the flatform, open-source Elasticsearch and Kibana are used as visualization tools. The data sent to the Elasticsearch database via Ni-Fi and then visualized in real-time with the Kibana interface. The Elasticsearch installed on the cluster allows you to search the data very quickly and to see the situations that occur within the desired date range via Kibana. As seen in Fig. 5, the bottom of the picture indicates the visualization of the selected pilot cell in Kibana. After the implementation of the pilot cell completed, the same process is applied to 10 injection machines (top of the Fig. 5) The outputs machine learning and advanced analytics applications which are built by using spark sent to Elasticsearch and visualized in Kibana to show valuable data to users.

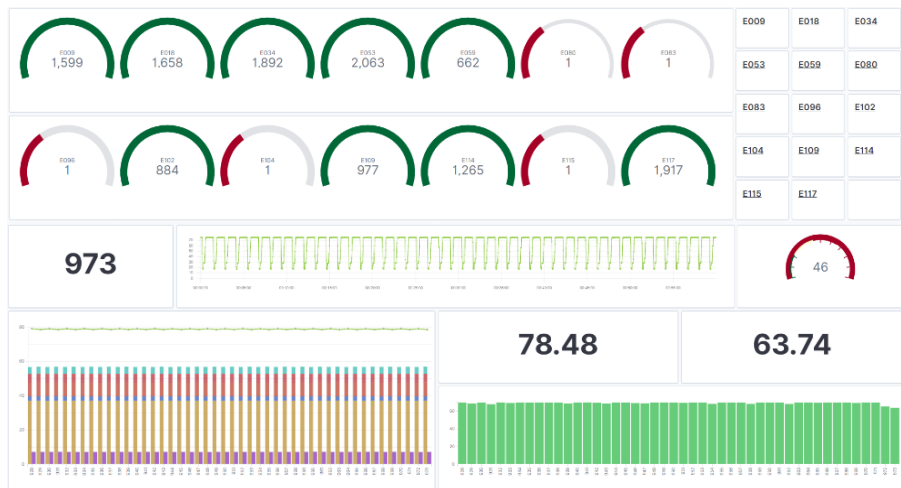

**Fig. 5.** Every second, 400 variables are collected from 10 injection cell (top), and pilot cell dashboard in Kibana.

### 1.6 Conclusion

In this paper, we describe the open-source big data ecosystem created in the Farplas manufacturing company, which has a great benefit in efficiency. This system, which can provide instant data flow from many systems, provides real-time on-premise data acquisition. In this way, the problems can be observed instantly, and in some cases, even a warning can be given before the problem occurs.

First, in the production layer data parameters such as machine name and model are defined in the main PLC via OPC software and published using the MQTT protocol. This data is then transferred to the Flatform big data platform. Apache Ni-Fi is used to transfer published data from production layer to the Flatform and create data pipelines. Then, data converted into JSON format on Ni-Fi is distributed to Elasticsearch, HDFS and other platforms with the help of Kafka. Spark and Kafka could be good choices in terms of the real-time data streaming and transmitting. All this installation environment is coordinated to work together, like a Hadoop eco-system.

Analytics Layer of Flatform contains Jupiter Notebook and Zeppelin to manage Spark and Python code to build artificial intelligence models. First predictive use-case model is a predictive quality model. The output of the model is visualized in Kibana by sending it to the Elasticsearch database via Spark as often as desired. In this way, machine operators can see the cause of the error instantly and change the parameters of the machine and prevent the production of more defective parts in the future.

In Farplas, using this big data infrastructure, abnormality in production are determined with the help of a cosine matrix of the energy values consumed by the machines. Then, new parameters are obtained with optimization studies. With the help of new machine parameters obtained with this method, energy consumption is reduced in production.

Future research activities would tend to increase artificial intelligence applications on this platform. AI models would be established with more parameters and solution methods. Besides, the number of machines connected to the system can also be increased in the next stage and connecting more system such as ERP in this platform.